\documentstyle[12pt,epsf]{article}
\newcommand{\abstracts}[1]{{
\centering{\begin{minipage}{12.2truecm}
\normalsize\baselineskip=15pt
\centerline{\footnotesize ABSTRACT}\vspace*{0.3cm}
\parindent=20pt #1
\end{minipage}}\par}}
\newcommand\Appendix[1]{\par
\setcounter{section}{0}
 \setcounter{equation}{0}
 \renewcommand{\thesection}{Appendix \Alph{section}}
\section{#1}
 \def\theequation{\Alph{section}.\arabic{equation}}}

\newcommand{\Z}{{Z \!\!\! Z}}
\newcommand{\dual}{\mbox{}^{\ast}}
\newcommand{\LL}{{I\!\! L}}

\newcommand{\eq}[1]{(\ref{#1})}

\newcommand{\beq}{\begin{equation}}
\newcommand{\eeq}{\end{equation}}
\newcommand{\beqn}{\begin{eqnarray}}
\newcommand{\eeqn}{\end{eqnarray}}

\newcommand{\cD}{{\cal D}}

\def\cC{{\cal C}}

\newcommand{\cZ}{{\cal Z}}
\newcommand{\cS}{{\cal S}}

\newcommand{\intpiD}{\int\limits_{-\pi}^{+\pi} {\cD}}

\newcommand{\intinfD}{\int\limits_{-\infty}^{+\infty} {\cD}}
\newcommand{\expb}[1]{\exp\left\{ #1 \right\} }
\newcommand{\CK}[1]{\mbox{\scriptsize C}_{\mbox{$\scriptstyle #1$}}}
\newcommand{\nsum}[2]{\sum_{ #1(\CK{#2}) \in \Z }}

\newcommand{\nnsum}[2]{\!\!\!\! \sum_{\dual #1(\dual\CK{#2})
\in \Z} \!\!\!\! }
\newcommand{\nddsum}[2]{\sum_{\stackrel{\scriptstyle \dual #1(\dual\CK{#2})
\in \Z} {\delta \dual #1=0}}}
\newcommand{\dd}{\mbox{d}}

\newcommand{\AmS}{{\protect\the\textfont2
 A\kern-.1667em\lower.5ex\hbox{M}\kern-.125emS}}

\def\dd{{\rm d}}

\def\NP{ Nucl.~Phys.}
\def\PL{ Phys.~Lett.}

\def\PRL{ Phys.~Rev.~Lett.}

\begin{document}
\begin{center}
\vspace{-1cm}
\begin{flushright}
{\large KANAZAWA-97-08}
\end{flushright}
\vspace{1.5cm}
{\baselineskip=24pt
{\Large \bf Aharonov--Bohm Effect in Lattice Abelian Higgs Theory}}\\

{\baselineskip=24pt

\vspace{1cm}

{\large M.N.~Chernodub$^{a,b}$, F.V.~Gubarev$^a$ and
M.I.~Polikarpov$^{a,b}$}\\

\vspace{.5cm}
{ \it

\vspace{0.3cm}

$^a$ ITEP, B.Cheremushkinskaya 25, Moscow, 117259, Russia

\vspace{0.3cm}

$^b$ Department of Physics, Kanazawa University,\\
Kanazawa 920-11, Japan

}
}
\end{center}

\vspace{1.5cm}

\abstracts{We study a field--theoretical analogue of the
Aharonov--Bohm effect in two-, three-  and  four-dimensional Abelian
Higgs models; the corresponding topological interaction is
proportional to the linking number of the Abrikosov vortex and the
particle world trajectories. We show that the Aharonov--Bohm effect
gives rise to a nontrivial interaction of charged test particles.
The numerical calculations in the three-dimensional model confirm
this fact.}


\section{Introduction}
\baselineskip=14pt

It is well known that the Abelian Higgs model in three and four
dimensions has  classical solutions called the
Abrikosov--Nielsen--Olesen strings (vortices)~\cite{AbNiOl}. These
strings carry quantized magnetic flux (vorticity),  and the wave
function of the charged particle which is scattered on the string
acquires an additional phase. The shift in the phase is a physical
effect which is the field-theoretical analogue  \cite{ABE} of the
quantum-mechanical Aharonov--Bohm effect \cite{AhBo59}: strings play
the role of solenoids  that  scatter the charged particles.
The topological long-range Aharonov--Bohm interaction between strings
and charged particles was discussed in the continuum theory
\cite{ABE} and on the lattice \cite{PoWiZu93}.

In Section~2 we show that in the lattice Abelian Higgs model with
non-compact gauge field  the Aharonov--Bohm effect gives rise to
long-range Coulomb-like interaction between test particles. In the
three-dimensional case the induced potential is confining, since the
Coulomb interaction grows logarithmically.
In Section~3 we give  the results of numerical calculations of
the induced potential. These results confirm the existence of
the Aharonov--Bohm effect in the Abelian Higgs model.

\section{Potential Induced by the Aharonov--Bohm Effect}

The partition function of  the  $D$-dimensional non-compact lattice
Abelian Higgs model~is\footnote{We  use  the notation  of the
calculus of differential forms on the lattice \cite{BeJo82}; see
Appendix A.}:

\beqn
  \cZ = \intinfD A \intpiD \varphi \nsum{l}{1} \expb{ - S(A,\varphi,l)}
  \,,\label{NC}
\eeqn
where

\beqn
  S(A,\varphi,l) = \beta \|\dd A\|^2 +
  \gamma \| \dd\varphi + 2\pi l - N A \|^2\,,
  \label{ActionNC}
\eeqn
$A$ is the non-compact gauge field, $\varphi$ is the phase of the
Higgs field and $l$ is the integer-valued one-form. For simplicity,
we consider the limit of infinite Higgs boson mass, the
radial part  $| \Phi |$  of the Higgs field, $|\Phi| e^{i\varphi}$
 being frozen. For the interaction of the Higgs field with the gauge
field we use the Villain form of the action.

One can rewrite  \cite{PoWiZu93} the partition function \eq{NC} as  a
sum over closed vortex trajectories ($D =3$) or as a  sum over
closed string world sheets ($D = 4$),  using the analogue of
Berezinski--Kosterlitz--Thauless (BKT) transformation
\cite{BKT}:

\beq
  \cZ \propto \cZ^{BKT} =  \hspace{-2mm} \nddsum{j}{2}
  \hspace{-2mm} \expb{ - 4 \pi^2 \gamma
  \left(\dual j, {(\Delta + m^2)}^{-1} \dual j \right)}
  \,, \label{TD}
\eeq
where $m^2 = N^2 \gamma \slash \beta$ is the $classical$ mass of the
vector boson $A$. The closed objects $\dual j$  defined on
the dual lattice have  dimension 1 and correspond to a closed vortex
for $D = 3$,  whereas  for $D = 4$ they have
dimension 2 and correspond to a closed string world sheet. These
objects $\dual j$, which are topological defects on the lattice,
interact with each other through the Yukawa forces ${(\Delta +
m^2)}^{-1}$.

Applying the same transformation to the quantum average of the Wilson
loop for the test particle of charge $M$, $W_M(\cC) = \exp\{ - i M
(A,j_\cC) \}$, we get the following formula~\cite{PoWiZu93}:

\beqn
 {<W_M(\cC)>}_N = \frac{1}{\cZ^{BKT}} \nddsum{j}{2}
 \exp\Bigl\{- 4\pi^2\gamma \Bigl(\dual j,{\Bigl(\Delta +
 m^2 \Bigr)}^{-1}\dual j \Bigr) \nonumber\\
 - \frac{M^2}{4 \beta}( j_\cC,(\Delta + m^2)^{-1} j_\cC) - 2 \pi
 i \frac{M}{N} \Bigl(\dual j_\cC,{\Bigl(\Delta + m^2\Bigr)}^{-1} \dd
 \dual j \Bigr) + 2 \pi i \frac{M}{N}
 \LL\left(j,j_\cC\right)\Bigr\}\,.
 \label{wl}
\eeqn
The first three terms in this expression are short-range Yukawa
forces between defects (strings or vortices) and test particles of
charge $M$. The last long-range term has  a
topological origin: $\LL (j,j_\cC)$ is the linking number
between the world trajectories of the defects $\dual j$ and the
world trajectory of the test particle~$j_\cC$:

\beq
       \LL(j ,j_\cC) = (\dual j_\cC, {\Delta}^{-1} \dd
       \dual j)\,.  \label{Ll}
\eeq
In three (four) dimensions,  the trajectory of the vortex (string)
$\dual j$ is a closed loop (surface) and the linking number $\LL$ is
equal to the number of points at which the loop $j_\cC$ intersects the
two (three) dimensional volume bounded by the loop (surface)~$\dual
j$.  Equation \eq{Ll} is the lattice analogue of the Gauss formula
for the linking number. This topological interaction corresponds to
the Aharonov--Bohm effect in field
theory~\cite{ABE,PoWiZu93}, the defects (vortices or strings) which
carry the magnetic flux ${2 \pi}/{N}$ scatter the test particle
of  charge $M$.

In the limit $N^2 \gamma \gg \beta$,  the partition function
\eq{TD} becomes

\beqn
  \cZ^{BKT}_0 = \nddsum{j}{2}
  \expb{- \frac{4 \pi^2 \beta}{N^2}
  {||\dual j||}^2 }\,.
 \label{TD2}
\eeqn
In the corresponding 3D (4D) continuum theory the term ${||\dual
j||}^2$ is proportional to the length (area) of the world trajectory
$\dual j$;  therefore,  the vortices (strings) are free. In the
said  limit,   the expectation
value of the Wilson loop $W_M(\cC)$ \eq{wl} is

\beqn
 {<W_M(\cC)>}_N = \frac{1}{\cZ^{BKT}_0} \nddsum{j}{2}
 \exp\Bigl\{- \frac{4 \pi^2 \beta}{N^2} {||\dual j||}^2 +
 2 \pi i \frac{M}{N} \LL(j,j_\cC) \Bigr\}\,.
 \label{wl2}
\eeqn
This formula describes the Aharonov--Bohm interaction of free
defects carrying magnetic flux ${2 \pi}/{N}$ with the test
particle of  charge $M$.

For $D=2$, the expression \eq{wl2} can be computed exactly. In
this case, the defects are pointlike, $\dual j$ is attached to sites
($\dual j = \dual j(\dual C_2)$) and the condition $\delta \dual j
=0$ is satisfied for any $\dual j$;  therefore,  all  the  variables
$\dual j$ in equations  (\ref{TD2}) and (\ref{wl2}) are independent.
The linking number can be written as $\LL(j,j_\cC) = (j, m_\cC)$,
where the two-form $m_\cC$ represents the surface spanned on
the contour $\cC$. After these remarks the evaluation of $W_M(\cC)$
becomes trivial and the result is as follows:

\beqn
 <W_M(\cC)> & =
 & {\left( \frac{ \sum\limits_{j \in \Z} \exp\left\{- \frac{4 \pi^2
 \beta}{N^2} j^2 + 2 \pi i \frac{M}{N} j \right\}}{ \sum\limits_{j
 \in \Z} \exp\left\{- \frac{4 \pi^2 \beta}{N^2} j^2 \right\}}
 \right)}^{\cS(\cC)} \nonumber\\ & = &  const. \exp\Bigl\{ -
 \sigma(M, N;\beta) \cdot \cS(\cC) \Bigr\}\,, \label{D2W}
\eeqn
where $\cS(\cC)$ is the area of the surface bounded by the contour
$\cC$. The string tension is

\beqn
 \sigma(M, N;\beta) = - ln\left[
 \frac{\Theta\left( \frac{4 \pi^2\beta}{N^2}, \frac{M}{N}
 \right)}{\Theta\left( \frac{4 \pi^2\beta}{N^2},0\right)}\right]\,,
 \label{sigma1}
\eeqn
where the $\Theta$--function has the form $\Theta(x,q) = \sum_{j \in
\Z} \exp\left\{ - x j^2 + 2 \pi i q j \right\}$. Note   that if the
charge of the test particle is completely screened by the charge of
the Higgs condensate $\left( M/N \in \Z \right)$, then  the string
tension is equal to zero, since $\Theta(x,q + n) = \Theta(x,q)$ ($n$
is an integer).  The area law of the Wilson loop for fractionally
charged particles in the $2D$ Abelian Higgs model was first found in
\cite{CaDaGr}, but then  it was not realized that the nonzero value
of the string tension is due to the analogue of the Aharonov--Bohm
effect.

For $D=3,4$, the  expression \eq{wl2} can be estimated in the
saddle--point approximation~\cite{Trento}.  Let us represent the
condition $\delta \dual j = 0$ in \eq{wl2} by introducing
integration over  an additional field $C$:

\beqn
  \nddsum{j}{2} \cdots = \nnsum{j}{2} \delta \left( \delta \dual j
  \right) \cdots = \intinfD \dual C \nnsum{j}{2}\, \exp
  \left\{ i (\delta \dual j, \dual C) \right\} \cdots \,.
  \label{A8}
\eeqn
Inserting unity
$1 = \intinfD \dual F \, \delta \left( \dual F - \dual j \right)$
into equation \eq{wl2}, we get

\beqn
  <W_M(\cC)> = \frac{1}{\cZ^{BKT}} \intinfD \dual C \intinfD \dual F
  \nnsum{j}{2} \cdot \nonumber\\
  \exp \left\{ - \frac{4 \pi^2\beta}{N^2} {\| \dual F \|}^2
  + 2 \pi i \frac{M}{N} (\dual m_\cC,\dual F) + i (\delta \dual F, \dual C)
  \right\} \cdot \delta \left( \dual F - \dual j \right) \,.
  \label{A9}
\eeqn
The use of the Poisson formula $\sum_{n \in \Z} \delta(n-x) = \sum_{2
\pi n \in \Z} \, e^{i n x}$ and integration over the fields
$\dual F$ leads us to the dual representation of the quantum average
\eq{wl2}:

\beqn
 <W_M(\cC)> = \frac{1}{\cZ^{BKT}} \intinfD \dual C
  \nnsum{j}{2} \exp \Bigl\{ - \frac{N^2}{16 \pi^2 \beta} {\| \dd
  \dual C + 2 \pi \frac{M}{N} \dual m_\cC + 2 \pi \dual j\|}^2
  \Bigr\}\,.
\label{A10}
\eeqn
For

\beqn
  N^2 \Delta^{-1}_{(D)}(0)\gg 4 \beta\,, \label{Condition}
\eeqn
where $\Delta^{-1}_{(D)}(R)$ is the $D$--dimensional massless lattice
propagator, we can evaluate this expression in the semiclassical
approximation: the result is

\beqn
  {<W_M(\cC)>}_N =
  const. \exp \Bigl\{ - \kappa^{(0)} (M, N;\beta) \cdot \left(j_\cC,
  \Delta^{-1} j_\cC\right)\Bigr\}\,., \label{wl3}
\eeqn
where

\beqn
  \kappa^{(0)} (M, N;\beta) = \frac{q^2 N^2}{4 \beta}\,, \qquad q =
  \min_{K \in \Z} |\frac{M}{N} - K|\,.
\label{q}
\eeqn
Here $q$ is the distance between the ratio $M \slash N$ and  the
nearest integer number. Just as equation \eq{sigma1}, the
expressions (\ref{wl3}), (\ref{q}) depend on the fractional part of
$M \slash N$, which  is a  consequence of the Aharonov--Bohm effect.
The interaction of the test charges is absent if $q = 0$ ($M \slash
N$ is integer);  this corresponds to the complete screening of the
test charge $M$ by the Higgs bosons of charge $N$.

To find the potential induced by the Aharonov--Bohm effect we
consider the product of two Polyakov lines: $W_M(\cC) = L^+_M(0)
\cdot L_M(R)$. Then $(j_\cC, \Delta^{-1} j_\cC) = 2 T \,
(\Delta^{-1}_{(D-1)}(R) + \Delta^{-1}_{(D-1)}(0))$ and  equation
\eq{wl3} is reduced to

\beqn
  {<L^+_M(0) \, L_M(R)>}_N =
  const. \, \exp \left\{ - 2 \, \kappa^{(0)}(M, N; \beta) \,
  T \, \Delta^{-1}_{(D-1)}(R)\right\}.
  \label{wlR}
\eeqn
For  large $R$,  we have \vspace {2pt} $\Delta^{-1}_{(2)}(R) =
\frac{c_3}{2} \ln R + \dots$, and $\Delta^{-1}_{(3)}(R) =
\frac{c_4}{2} R^{-1} + \dots$, where $c_3$ and $c_4$ are  constants.
Thus, the Aharonov--Bohm effect in three- and four-dimensional
Abelian Higgs model gives rise in the leading order to the following
long-range potentials:

\beqn
  V^{(0)}_{3D}(R) = 2 \, \kappa^{(0)}(M, N; \beta) \,
  T \, \Delta^{-1}_{(2)}(R) =
  c_3 \, \kappa^{(0)}(M, N; \beta) \cdot \ln R
  + O\left( R^{-1} \right)\,,\label{V3}\\
  V^{(0)}_{4D}(R) = 2 \, \kappa^{(0)}(M, N; \beta) \,
  T \, \Delta^{-1}_{(3)}(R) =
  c_4 \, \kappa^{(0)}(M, N; \beta) \cdot \frac{1}{R}
  + O\left( R^{-2} \right)\,.\label{V4}
\eeqn

A $3D$ vortex model in the continuum is discussed in \cite{Sa79},
where  it is  shown that the potential has the form $V^c_{3D}(R)
\approx const.\, q^2 \psi_0 \, \ln \frac{R}{R_0}$, with $\psi_0$
being  proportional to the vortex condensate, and $R_0$ having the
order of the vortex width. This result is in agreement with our
estimation \eq{V3} of the potential.

Note that from expression \eq{wl2} the following properties of
the potential $V_{(M,N)}$ can be obtained:

\beqn
  V_{(M,N)} = V_{(N - M,N)}\,, \qquad V_{(N,N)} = 0\, .
  \label{gen}
\eeqn
The long-range potentials $V(R)$,  equations (\ref{V3}), (\ref{V4}),
and the string tension \eq{sigma1} satisfy these relations.

\section{Numerical Calculations}

We calculated numerically the potential between the test particles
of charge $M$ in the three-dimensional Abelian Higgs model,
the Higgs boson charge being  $N=6$. The action of the model is chosen in
the Wilson form:  $S[A,\varphi] = \beta {||\dd A||}^2 - \gamma \cos
(\dd \varphi + N A)$. We perform the calculations by the standard
Monte--Carlo method and we use 200 statistically independent
configurations for each point in the  $\beta-\gamma$ plain. The
simulations are performed on lattice of size $16^3$ for the
test charges $M = 1, \dots, N$.

We fit the numerical data for the Wilson loop quantum average
by the following formula:

\beqn
- \ln {<W_M (C)>}_N = \kappa_{num} (j_\cC, \Delta^{-1} j_\cC) +
m^2_{num} {||j_\cC||}^2 + C_{num}\,, \label{fit}
\eeqn
where $\kappa_{num}$, $m_{num}$ and $C_{num}$ are the fitting
parameters. It turns out that the expectation values of the Wilson
loops of the shapes $L_1 \times L_2$, $L_1,L_2=1,\dots,5$ are well
described by~\eq{fit}.

\begin{figure}[bth]
\centerline{\epsfxsize=0.9\textwidth\epsfbox{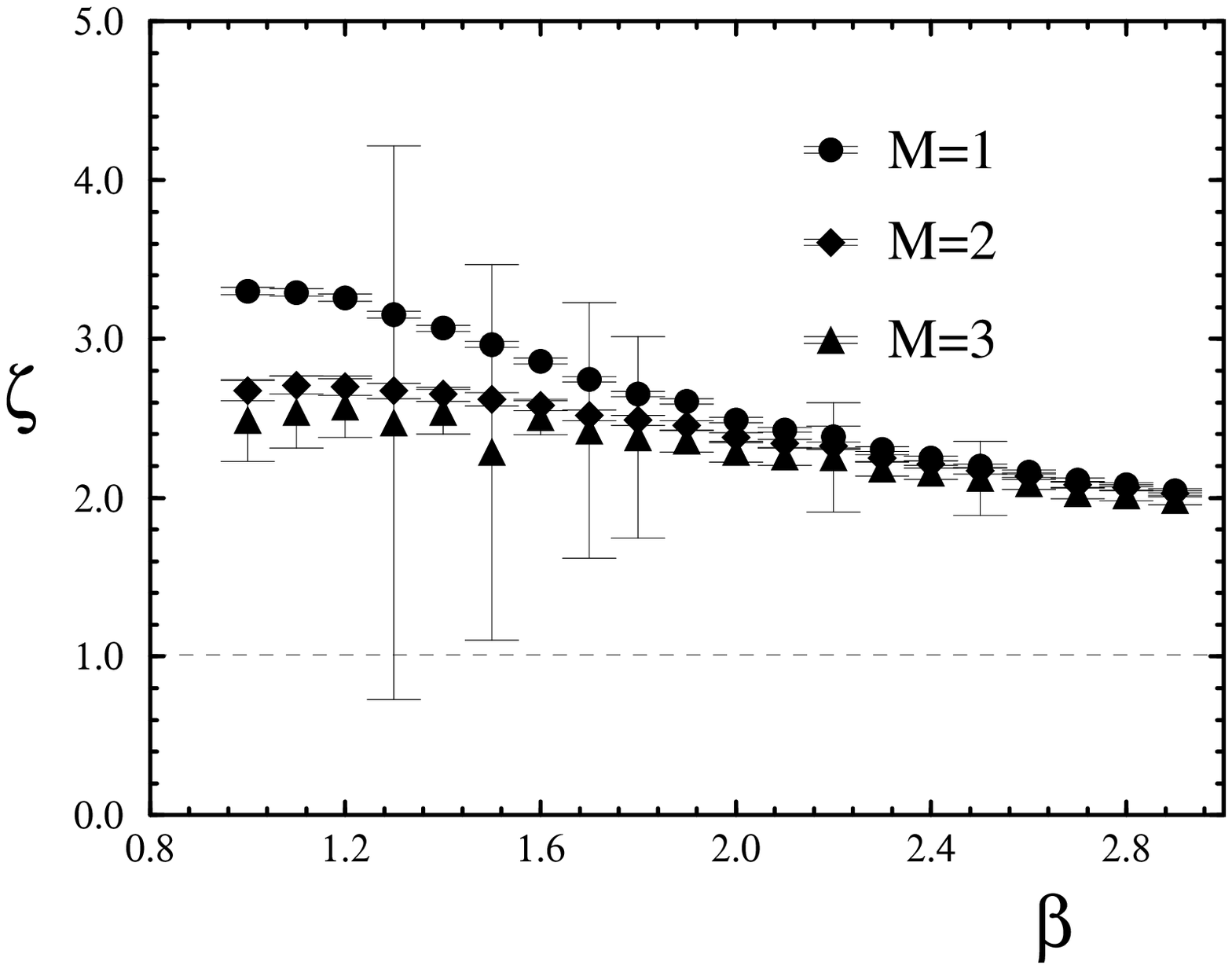}}
\vspace{0.3cm}
\centerline{Fig.~1: The ratio $\zeta=
\frac{\kappa_{num}}{\kappa^{(0)}}$ $vs.$ $\beta$ for $M=1,2,3$ at
$\gamma=1$.}
~
\end{figure}

In Fig.~1 we show the dependence of the
coefficients $\zeta=\kappa_{num} \slash \kappa^{(0)}$ on $\beta$ for
$M=1,2,3$ and $\gamma = 1$, where $\kappa^{(0)}$ is given by
the semiclassical expression \eq{q}. We studied the region $1\le
\beta \le 3$ for which the condition \eq{Condition} of validity of
the semiclassical expression \eq{q} is not valid\footnote{Condition
\eq{Condition} for $N=6$ reads as follows: $\beta \ll 2.2$.
Unfortunately, even at $\beta < 1$ our data for $\zeta$ contains
large error bars, therefore we do not show this region of $\beta$ on
the figure.} (note that $\Delta^{-1}_{(3)} (0)  
\approx 0.24$.).  The considered values of the parameters $\beta$ and 
$\gamma$ correspond to the Coulomb phase in which the vortices are 
condensed.

It is clearly seen from Fig.~1 that for $\beta > 2.3$ the values  of
$\zeta$ are independent of $M$.  Thus,  $\kappa_{num}$ is
proportional to $\kappa^{(0)}$ with a coefficient which differs
from unity because of quantum corrections. The same effect is
observed in this model at  finite temperature~\cite{GuPoCh96}.

The numerical data also shows that $\kappa^{num}_{M} =
\kappa^{num}_{N - M}$ and the potential for the charge $M=N$ is equal
to zero within numerical errors for all considered values of $\beta$.
These relations are in agreement with the Aharonov--Bohm nature of
the potential $V_{(M,N)}(R)$ (cf. equation \eq{gen}).  Note that the
usual Coulomb interaction of the test particles of charge $M$ is
proportional to $M^2$.  Therefore, the fact that the potential
$V_{(M)}$ satisfies the relations \eq{gen} means that the usual
Coulomb interaction of the test particles is small.


\section*{Conclusion and Acknowledgments}


We have shown by analytical calculations that the Aharonov--Bohm
effect induces the long-range interaction (\ref{V3},\ref{q}) at small
values of $\beta$: $\beta \ll N^2 \gamma$, $\beta \ll
\Delta^{-1}_{(3)} (0) N^2 \slash 4$. Our numerical simulations for
$\beta \ge 2.3$ show that this effect leads to long-range potential
which is proportional to potential \eq{V3}: $V_{3D}(R;M,N,\beta) =
\zeta(\beta) V^{(0)}_{3D}(R;M,N,\beta)$, where $\zeta$ is independent
of $M$. At the intermediate values of $\beta$ the Aharonov--Bohm
effect also induces a long-range potential but the induced potential
is not proportional to the semiclassical expression eq.\eq{V3}.
The potential depends non-analytically on the charge of the test
particle. Due to the long-range nature of the induced potential,
the Aharonov--Bohm effect may be important for the dynamics of the
colour confinement in nonabelian gauge theories~\cite{ChPoZu94}.

M.N.Ch. and M.I.P. acknowledge the kind hospitality of the
Theoretical Department of the  Kanazawa University. The authors are
grateful to  D.A.~Ozerov for useful discussions. This work is
supported by the JSPS Program on Japan -- FSU scientists
collaboration, and also by the Grants:  INTAS-94-0840, INTAS-94-2851,
INTAS-RFBR-95-0681,  and  Grant No. 96-02-17230a  of  the Russian
Foundation for Fundamental Sciences.

\newpage

\Appendix{ }

Let us briefly summarize the main notions from the theory of
differential forms on the lattice \cite{BeJo82}.  The advantage of the
calculus of differential forms consists in the general character of
the expressions obtained. Most of the transformations depend neither
on the space-time dimension  nor on the rank of the fields. With
minor modifications, the transformations are valid for lattices of any
form (triangular, hypercubic, random, {\it etc}). A differential form
of rank $k$ on the lattice is a function $\phi_{k}$ defined on
$k$-dimensional cells $C_k$ of the lattice, {\it e.g.}, the scalar
(gauge) field is a 0--form (1--form). The exterior differential
operator {\it d} is defined as follows:

\beq
(\dd \phi ) (C_{k+1}) =\sum_{\CK{k} \in \partial\CK{k+1}} \phi(C_{k}).
\label{def-dd}
\eeq
Here $\partial C_{k}$ is the oriented boundary of the $k$-cell
$C_{k}$. Thus,  the operator {\it d} increases the rank of the form
by unity; $\dd \varphi$ is the link variable constructed, as usual,
in terms of the site angles $\varphi$, and $\dd A$ is the plaquette
variable constructed from the link variables $A$.  The scalar product
is defined in the standard way:  if $\varphi$ and $\psi$ are
$k$-forms, then $(\varphi,\psi)=\sum_{C_k}\varphi(C_k)\psi(C_k)$,
where $\sum_{C_k}$ is the sum over all cells $C_k$.
To any $k$--form on the $D$--dimensional lattice there
corresponds a $(D-k)$--form $\dual\Phi(\dual C_k)$ on the dual
lattice, $\dual C_k$ being the $(D-k)$--dimensional cell on the dual
lattice. The co-differential $\delta=\dual \dd \dual$ satisfies the
partial integration rule  $(\varphi,\delta\psi)=(\dd\varphi,\psi)$.
Note that $\delta \Phi(C_k)$ is a $(k-1)$--form and
$\delta \Phi(C_0) = 0$. The norm is defined by  $\|a\|^2=(a,a)$.
Therefore, $\| \dd\varphi + 2\pi l - N A \|^2$ in \eq{ActionNC}
implies summation over all links.  The sum $\nsum{l}{1}$ is taken
over all configurations of the integers $l$ attached to the links
$C_1$. Due to the well-known property $\dd^2 = \delta^2 = 0$, the
action \eq{ActionNC} is invariant under the gauge transformations $A'
= A + \dd \alpha$, $\varphi' = \varphi + N \alpha$ . The lattice
Laplacian is defined by  $\Delta = \dd\delta + \delta\dd$.

\end{document}